\documentclass[aps,pra,twocolumn,showpacs,superscriptaddress]{revtex4-1}
\usepackage{amsfonts}
\usepackage{graphicx}
\usepackage{hyperref}
\usepackage{amsmath}
\usepackage{amssymb}
\usepackage[usenames, dvipsnames]{color}
\usepackage{subfigure}
\usepackage{lipsum}

\pdfoutput=1

\begin{document}

\title{Reduction of Frequency-dependent Light Shifts in Light-narrowing
Regimes: A Study Using Effective Master Equations }
\author{Yue Chang}
\email{yuechang7@gmail.com}
\author{Yu-Hao Guo}
\author{Jie Qin}
\email{jie.qin@yahoo.com}

\begin{abstract}
Alkali-metal-vapor magnetometers, using coherent precession of polarized
atomic spins for magnetic field measurement, have become one of the most
sensitive magnetic field detectors. Their application areas range from
practical uses such as detections of NMR signals to fundamental physics
research such as searches for permanent electric dipole moments. One of the
main noise sources of atomic magnetometers comes from the light shift that
depends on the frequency of the pump laser. In this work, we theoretically
study the light shift, taking into account the relaxation due to the optical
pumping and the collision between alkali atoms and between alkali atoms and
the buffer gas. Starting from a full master equation containing both the
ground and excited states, we adiabatically eliminate the excited states and
obtain an effective master equation in the ground-state subspace that shows
an intuitive picture and dramatically accelerates the numerical simulation.
Solving this effective master equation, we find that in the light-narrowing
regime, where the line width is reduced while the coherent precession signal
is enhanced, the frequency-dependence of the light shift is largely reduced,
which agrees with experimental observations in cesium magnetometers. Since
this effective master equation is general and is easily solved, it can be
applied to an extensive parameter regime, and also to study other physical
problems in alkali-metal-vapor magnetometers, such as heading errors.
\end{abstract}

\maketitle

\affiliation{Beijing Automation Control Equipment Institute, Beijing 100074,
China}
\affiliation{Quantum Technology R$\&$D Center of China Aerospace
Science and Industry Corporation, Beijing 100074, China}



\section{Introduction}

Alkali-metal-vapor atomic magnetometers \cite%
{Kominis2003,Budker2003,Budker2007}, which have become one of the most
sensitive devices for magnetic field detection, find applications in various
areas ranging from practical uses such as NMR signal detection \cite%
{Budker2007,PhysRevLett.93.160801,PhysRevLett.94.123001,Walker1997} to
fundamental physics research such as searches for permanent electric dipole
moments \cite{fortson2003search,PhysRevA.75.063416,roberts2015}. The physics
behind atomic magnetometers is as follows: polarized atomic spins precess
along the magnetic field to be measured, and its precession angle, or the
so-called Larmor frequency that can be measured, is proportional to the
magnitude of the magnetic field. To have a collective spin precession for
measurement, the electronic spins are polarized by optical pumping \cite%
{RevModPhys.44.169,Happer1987,happer2010optically,auzinsh2010optically}.
However, in the measurement of the Larmor frequency, the light shift,
resulting from the interaction of light (the pump beam here) and matter,
behaves as an effective magnetic field to the atomic spins, and subsequently
shifts its precession frequency \cite%
{PhysRev.163.12,Mathur1968,PhysRevA.58.1412,seltzer2008developments,Schultze2017}%
. This light shift is dependent on the intensity and frequency of the pump
laser. Therefore, it will decrease the measurement accuracy if the pump
beam's frequency has fluctuations. One way to reduce this frequency
dependence of the light shift is to decrease the pump beam's intensity, or
to increase the line broadenings of the alkali atoms' excited states, but
both will lower the atomic polarization, which reduces the precession signal.

Recently, we have found that in cesium vapor magnetometers with buffer gas N$%
_{2}$ \cite{Guo:19}, without tuning the pump beam's intensity or the excited
states' lifetimes, the light shift's dependence on the laser frequency can
be greatly reduced in the light-narrowing regime \cite%
{PhysRevA.59.2078,PhysRevA.84.043416}, in which the line width of the spin
precession signal is narrowed and the fundamental sensitivity \cite%
{PhysRevA.84.043416,PhysRevA.94.013403}, which is inversely proportional to
the square root of the spin's transverse relaxation time, is improved, which
further improves the measurement accuracy.

\begin{figure}[tbp]
\begin{center}
\includegraphics[width=0.85\linewidth]{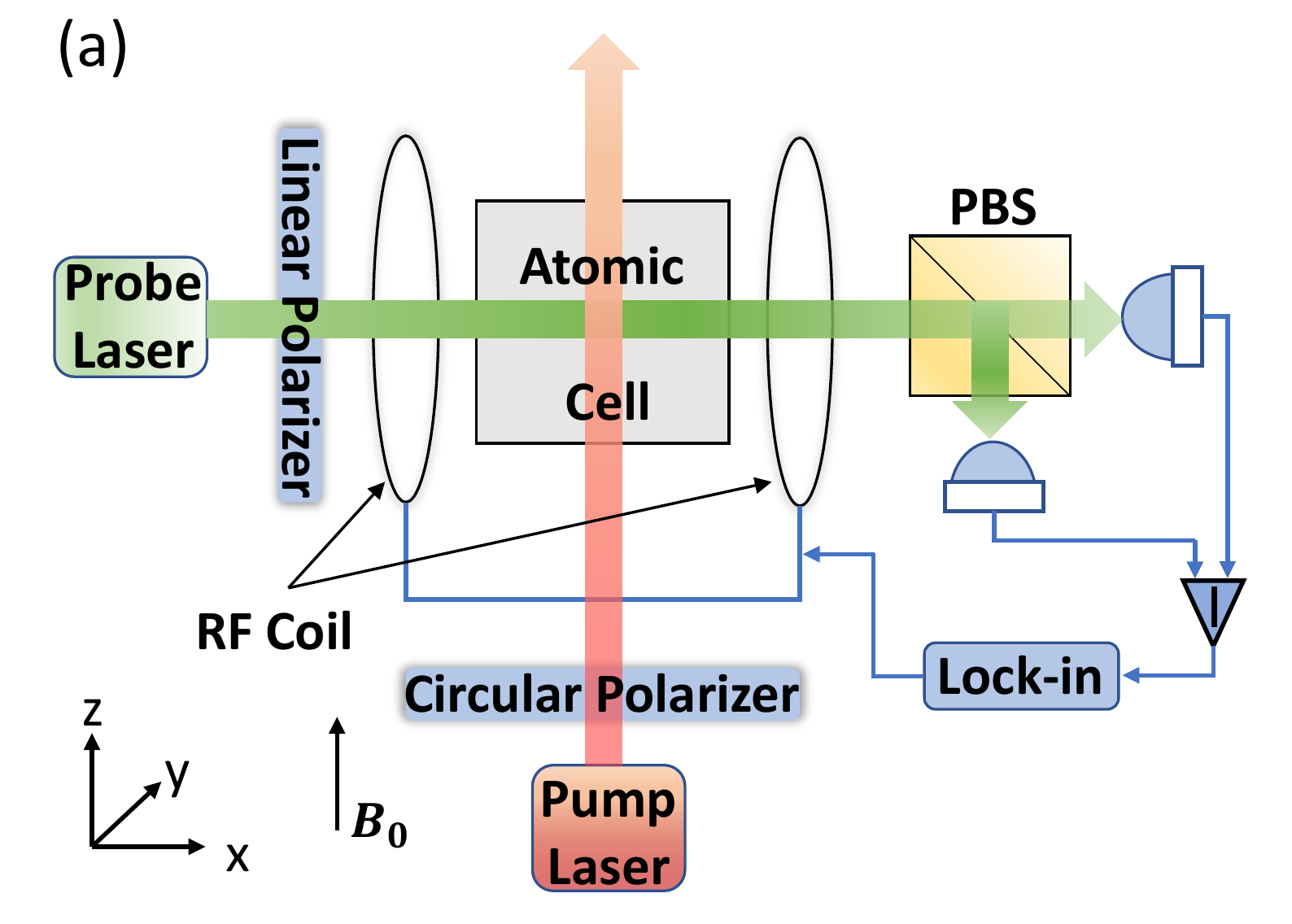}
\includegraphics[width=0.85\linewidth]{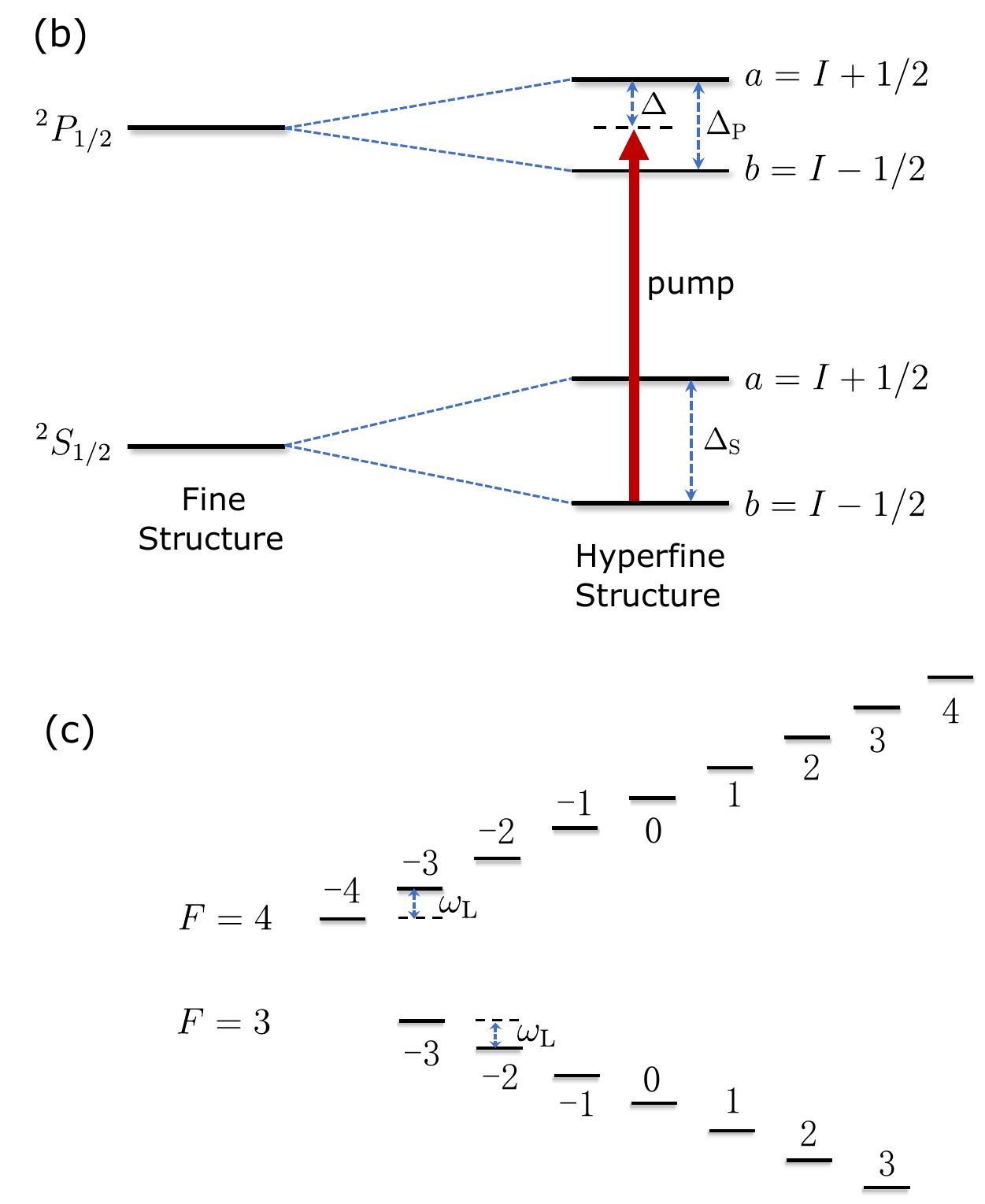}
\end{center}
\caption{(a) Experimental setup of the alkali-metal-vapor magnetometer. The
magnetic field $\vec{B}_{0}$ to be measured is in the $z$-direction, and a
pump laser (the red arrow), also propagating in the $z$-direction with
circular polarization, is applied to polarize the atomic spins to the $z$%
-direction. An oscillating magnetic field along the $x$-direction is
produced by two RF coils, and the resultant $x$-directional atomic spin
polarization is probed by a linearly polarized laser (the green arrow) along
the $x$-direction using its optical rotation. Note that the frequency of the
probe laser is about $80$ GHz blue detuned from the D2 transition, with the
laser power around $10$ mW, so it is a good approximation that the probe
laser is not taken into account in the optical pumping process. (b)
Schematic of the alkali atoms' D1 transition in the optical pumping process.
Here, $\left. ^{2}S_{1/2}\right. $ and $\left. ^{2}P_{1/2}\right. $
represent ground and first excited states, respectively, in the fine
structure, and through the hyperfine interaction between the electrons and
the nuclei, these levels are further split, with the splitting $\Delta _{%
\text{S}}$ ($\Delta _{\text{P}}$) for the $\left. ^{2}S_{1/2}\right. $ ($%
\left. ^{2}P_{1/2}\right. $) states. The pump laser (the red arrow) with
Rabi frequency $\Omega $ and detuning $\Delta $ (with respect to the
frequency difference between the $b=I-1/2$ ground and $a=I+1/2$ excited
states) induces transitions between the ground and first excited states. (c)
Ground state Zeeman sublevels for cesium ($I=5/2)$, with magnetic numbers
marked above/below each level. The energy difference between two adjacent
sublevels in the same multiplet is the Larmor frequency $\protect\omega _{%
\mathrm{L}}$. }
\label{fig1}
\end{figure}

In the experimental setup shown in Fig.~\ref{fig1}(a), an atomic cell
containing cesium atoms and nitrogen gas (buffer gas) is illuminated by a
circularly polarized pump laser propagating along the $z$-direction. The
magnetic field $\vec{B}_{0}$ to be measured is also in the $z$-direction,
and an oscillating magnetic field along the $x$-direction is generated by
two RF coils to induce atomic spin polarizations in the $x$-direction, which
are reconstructed by measuring the optical rotation of a linearly polarized
probe laser propagating in the $x$-direction. The energy levels of an alkali
atom are shown in Fig.~\ref{fig1}(b), where the electrons' fine structure
energy levels are denoted by $\left. ^{2}S_{1/2}\right. $ for the ground
states and $\left. ^{2}P_{1/2}\right. $ for the excited states. These fine
structure levels are further split by hyperfine interaction, with $\Delta _{%
\text{S}}$ ($\Delta _{\text{P}}$) the splitting between the two multiplets $%
F=a\equiv I+1/2$ and $F=b\equiv I-1/2$ states in the ground (first excited)
states. Here, only the D1 transition \cite{steck2003cesium} is under
consideration, since the pump laser is nearly resonant with the transition
frequency between the ground states and first excited states (the definition
of the detuning $\Delta $ is shown in Fig.~\ref{fig1}(b)), and the probe
laser is not taken into account in the optical pumping process since it is
far detuned from both the D1 and D2 transitions (about $80$ GHz blue detuned
from the D2 transition, with the laser power around $10$ mW). With the
magnetic field, the magnetic levels for cesium atoms are shown in Fig.~\ref%
{fig1}(c), with the Larmor frequency $\omega _{_{\mathrm{L}}}$. Note that
all the frequencies in this paper are the regular frequencies and not the
angular ones.

For the mechanism of this frequency-dependent light-shift reduction, an
intuitive picture is as follows. In the light-narrowing regime with $\Delta
=0$, the $b$\ ground states are pumped strongly and the alkali atoms mainly
populate the $a$ ground state, where most of the magnetic resonance is
generated and probed. Considering the Lorentzian form of the AC Stark shift
\cite{PhysRev.100.703} for a single state, one might conclude that the
dependence of the light shift on the pump beam's frequency is reduced
because of the large hyperfine splitting $\Delta _{\text{S}}$ in the ground
state (compared with the line width of the excited states). Note that we do
not choose the probe laser's frequency so that it only measures the magnetic
resonance from the $a$ multiplet. Actually, the only function of the probe
laser is to measure the response of the atomic spin to the oscillating
magnetic field. The fact that the states being pumped differ from the states
where most of the spin precession signal is generated arises naturally in
the light-narrowing regime with properly tuned pump laser powers.

However, the atomic ground states are incoherently coupled to each other by
the light-matter interaction and atomic collisions, so that the light shift
cannot be simply written as a Lorentzian or a sum of Lorentzians. Thus we
use the master equation to study the light shift in a general
alkali-metal-vapor atomic magnetometer, taking into account the light-matter
interaction and the relaxation due to collisions between alkali atoms and
between alkali atoms and buffer gas \cite{PhysRevA.58.1412}. The interaction
between the pump light and the alkali atoms is modeled using the dipole
approximation and rotating-wave approximation \cite{Walls2008,Gardiner2004}.
This master equation appeared in some early textbooks and papers \cite%
{RevModPhys.44.169,happer2010optically}, but it is not easily solved because
of its nonlinearity (caused by the mean field approximation for the
spin-exchange interaction) and its large superspace \cite%
{RevModPhys.44.169,happer2010optically,Gardiner2004}. (The full master
equation is in a Hilbert space consisting of all the ground and first
excited states.)\ Thus we adiabatically eliminate the excited states in the
weak-driving limit, where the Rabi frequency--the coupling strength between
the ground and excited states--is much smaller than the excited states'
decay rates, to acquire an effective master equation in a subspace
consisting of only the ground states. This can dramatically decrease the
calculation power and time needed to solve the nonlinear master equation,
and it explicitly shows the intuitive picture of the reduction of the
light-shift dependence on the frequency, as well as the light-narrowing
effect \cite{PhysRevA.59.2078}. With little cost of calculation, the light
shift and line width obtained by solving this effective master equation and
using linear response theory \cite{fetter2012quantum} agree well with the
experimental data.

The rest of this paper is organized as follows. In section \ref{s2}, we
model the system by a full master equation for the density matrix evolution
of the alkali atoms, including all the ground states and the first excited
states. Starting from this full master equation, in section \ref{s3}, we
adiabatically eliminate the excited states in the weak-driving limit and
obtain an effective master equation in only the ground-state subspace. It is
shown that this effective master equation can give the rate equations \cite%
{PhysRevA.60.3867} used in many contexts. And when the energy-level
broadening of the excited states is much larger than the hyperfine
splittings $\Delta _{\text{P}}$ and $\Delta _{\text{S}}$, the light-matter
interaction is reduced to a dissipation term that consists of only the
electronic spin operators \cite{PhysRevA.58.1412}, leading to the spin
temperature distribution. In section \ref{s4}, we study the linear response
of the alkali atoms to the small transverse oscillating magnetic field, both
analytically and numerically, showing good agreement between the theoretical
predictions and experimental data on both the light shift and line width in
a wide frequency regime of the pump laser. Finally, in section \ref{s5}, we
summarize our work and show other possible applications of the effective
master equation.

\section{Full master equation description}

\label{s2} In this section, we give the full master equation depicting the
time evolution of the density matrix $\rho \left( t\right) $ of the alkali
atoms. This master equation involves all the energy levels in the ground
state and the first excited states \cite%
{RevModPhys.44.169,happer2010optically}, which can be written as a sum of
four Lindblad operators,%
\begin{equation}
\partial _{t}\rho =\sum_{n=1}^{4}\mathcal{L}^{\left( n\right) }\rho ,
\label{me}
\end{equation}%
each coming from a different interaction. The first Lindblad term describes
the light-matter interaction. Without lost of generality, we assume the pump
laser is propagating parallel to the magnetic field's direction, which
defines the magnetic numbers of the hyperfine states, and is left-handed
circularly polarized. But this can be easily generalized to the opposite
case, i.e., a parallel propagating laser with right-handed circular
polarization. This will not change the conclusion of this paper. With the
left-handed circularly polarized pump laser, the light-matter interaction
contributes to the master equation as%
\begin{eqnarray}
\mathcal{L}^{\left( 1\right) }\rho  &=&-i\left[ H_{\text{lm}},\rho \right] +
\notag \\
&&\Gamma _{\text{sd}}\sum_{l=0,\pm 1}\left( \left\vert s\right\rangle
\left\langle p_{l}\right\vert \rho \left\vert p_{l}\right\rangle
\left\langle s\right\vert -\frac{1}{2}\left\{ \rho ,\left\vert
p_{l}\right\rangle \left\langle p_{l}\right\vert \right\} \right) ,
\end{eqnarray}%
where $\Gamma _{\text{sd}}$ is the spontaneous decay rate resulting from the
interaction between the alkali atoms and light in the free space; $%
\left\vert s\right\rangle $ and $\left\vert p_{l}\right\rangle $ are the
electron's orbital states $1s$ and $2p$, respectively; and $l$ in $%
\left\vert p_{l}\right\rangle $ is its quantum magnetic number. The
Hamiltonian $H_{\text{lm}}$ depicting the coupling between the pump beam and
the alkali atoms is written in the rotating frame with respect to the
laser's frequency as%
\begin{equation}
H_{\text{lm}}=\Omega ^{\prime }\left( \left\vert s\right\rangle \left\langle
p_{1}\right\vert +\left\vert p_{1}\right\rangle \left\langle s\right\vert
\right) ,
\end{equation}%
where $\Omega ^{\prime }$ is the Rabi frequency, and the dipole and
rotating-wave approximations \cite{Walls2008,Gardiner2004} are used. The
radiation trapping \cite{molisch1998radiation} effect is not included, since
the quenching \cite{seltzer2008developments} gas can largely remove it.

The second Lindblad operator depicts the alkali atoms' energy levels and
their Zeeman splitting due to the static magnetic field $\vec{B}_{0}=B_{z}%
\hat{e}_{z}$:%
\begin{equation}
\mathcal{L}^{\left( 2\right) }\rho =-i\left[ H_{\text{hf}}+H_{\text{Zee}%
},\rho \right],
\end{equation}%
where%
\begin{eqnarray}
H_{\text{hf}} &=&\sum_{M}\Delta _{\text{s}}\left\vert s_{\text{a}%
}M\right\rangle \left\langle s_{\text{a}}M\right\vert -\Delta _{\text{p}%
}\left\vert p_{\text{b}}M\right\rangle \left\langle p_{\text{b}}M\right\vert
\notag \\
&&+\Delta \sum_{\text{F=a,b}}\left\vert p_{\text{F}}M\right\rangle
\left\langle p_{\text{F}}M\right\vert
\end{eqnarray}%
gives the hyperfine structures and%
\begin{equation}
H_{\text{Zee}}=\sum_{\text{F=a,b}}\sum_{M}M\omega _{_{\text{F}}}\left\vert
s_{\text{F}}M\right\rangle \left\langle s_{\text{F}}M\right\vert
\end{equation}%
gives the Zeeman splitting. Here, $\left\vert s_{\text{F}}/p_{\text{F}%
}M\right\rangle $ is the hyperfine state in the $1s$ ($\left\vert s_{\text{F}%
}M\right\rangle $) or $1p$ ($\left\vert p_{\text{F}}M\right\rangle $, whose
energies have been shifted with respect to the pump beam's frequency)
orbital, with the total angular momentum $F=a$, $b$ and its projection in
the $z$-direction $M$. In $H_{\text{Zee}}$,$\ \omega _{_{\mathrm{a}%
}}=-\omega _{_{\mathrm{b}}}=\gamma _{\mathrm{e}}B_{z}/(2I+1)\equiv \omega
_{_{\mathrm{L}}}$ is the Larmor frequency of the atom, where $\gamma _{%
\mathrm{e}}$ is the gyromagnetic ratio of the electron. Note that only the
linear Zeeman splitting in the ground states has been considered, since
other interactions with the magnetic field, such as the nonlinear Zeeman
interaction for $B_{z}=0.1G$ and the Zeeman splitting in the excited states,
are too small to affect the result.

Since there are many alkali atoms and there is much buffer gas (nitrogen in
the experiment) in the heated atomic cell, collisions between atoms must be
taken into account, resulting in dissipation in the master equation as%
\begin{eqnarray}
\mathcal{L}^{\left( 3\right) }\rho &=&\gamma \left( \boldsymbol{S}\cdot \rho
\boldsymbol{S-}\frac{1}{2}\left\{ \rho ,\boldsymbol{S}\cdot \boldsymbol{S}%
\right\} \right)  \notag \\
&&+\frac{1}{2}\gamma _{\text{se}}\left\langle S_{z}\right\rangle \left(
S_{+}\rho S_{-}-S_{-}\rho S_{+}\boldsymbol{+}\left\{ \rho ,S_{z}\right\}
\right)  \notag \\
&&+\frac{1}{2}\gamma _{\text{se}}\left\langle S_{+}\right\rangle \left(
S_{-}\rho S_{z}-S_{z}\rho S_{-}\boldsymbol{+}\frac{1}{2}\left\{ \rho
,S_{-}\right\} \right) +\text{H.c.}  \notag \\
&&+\Gamma _{\text{pb}}\sum_{m=0,\pm 1}A_{m}\rho A_{j}^{\dagger }-\frac{1}{2}%
\left\{ \rho ,A_{m}^{\dagger }A_{m}\right\} ,
\end{eqnarray}%
where $\boldsymbol{S}$ is the electronic spin operator in the ground state, $%
S_{\pm }=S_{x}\pm iS_{y}$ is the spin raising/lowering operator, $\gamma _{%
\text{se}}$ is the spin exchange rate coming from collisions between alkali
atoms, and $\gamma =\gamma _{\text{se}}+\gamma _{\text{sd}}$ is the total
relaxation rate with the spin destruction rate $\gamma _{\text{sd}}$ coming
from collisions between alkali atoms and nitrogen molecules. In addition to
spin relaxation, the collisions also cause line broadening of the excited
states, with $\Gamma _{\text{pb}}$ being the pressure broadening of the $%
^{2}P_{1/2}$ states due to collisions of the alkali atoms with the nitrogen
molecules. During such a collision, the alkali atom in excited states decays
to the ground states by transferring its momentum to the nitrogen molecule's
angular momentum, rather than emitting photons. In $\mathcal{L}^{\left(
3\right) }$,\ the jump operators $A_{m}$ are defined as%
\begin{equation}
A_{0}=\sum_{m=\pm 1/2}\left\vert ^{2}S_{1/2},m\right\rangle \left\langle
^{2}P_{1/2},m\right\vert ,
\end{equation}%
\begin{equation}
A_{\pm 1}=\left\vert ^{2}S_{1/2},\mp \frac{1}{2}\right\rangle \left\langle
^{2}P_{1/2},\pm \frac{1}{2}\right\vert .
\end{equation}%
\ It can be shown straightforwardly from $\mathcal{L}^{\left( 3\right) }\rho
$ that the spin exchange interaction does not change the mean values of the
spins, i.e., $\partial _{t}^{\left( 3\right) }\left\langle \boldsymbol{S}%
\right\rangle =0$ if we set $\gamma _{\text{sd}}=0$ and $\Gamma _{\text{pb}%
}=0$, while the spin destruction interaction exponentially decreases the
spin's mean values, i.e., $\partial _{t}^{\left( 3\right) }\left\langle
\boldsymbol{S}\right\rangle =-\gamma _{\text{sd}}\left\langle \boldsymbol{S}%
\right\rangle $ if we set $\gamma _{\text{se}}=0$\ and $\Gamma _{\text{pb}%
}=0 $. Here, the time derivative $\partial _{t}^{\left( 3\right) }$ means we
consider only $\mathcal{L}^{\left( 3\right) }\rho $ in the time evolution of
the density matrix: $\partial _{t}^{\left( 3\right) }\rho =\mathcal{L}%
^{\left( 3\right) }\rho $.

To measure the precession frequency, a small oscillating magnetic field $%
B_{x}\hat{e}_{x}\cos \omega t$ along the $x$-direction, with amplitude $%
B_{x} $ and frequency $\omega $,\ is applied, leading to a time-dependent
term in the master equation,%
\begin{equation}
\mathcal{L}^{\left( 4\right) }\rho =-i\gamma _{\mathrm{e}}B_{x}\cos \omega t%
\left[ S_{x},\rho \right] .
\end{equation}

Note that the dimension of the superspace \cite%
{RevModPhys.44.169,happer2010optically,Gardiner2004} of the full master
equation is $4\left( 4I+2\right) ^{2}$, i.e., there are $4\left( 4I+2\right)
^{2}$ coupled nonlinear equations to be solved, hence the numerical
simulation consumes much time and power. In any case, the physics cannot be
revealed in such a big set of nonlinear equations. Therefore, we will
simplify this master equation by adiabatically eliminating the excited
states.

\section{Effective master equation in the ground-state subspace}

\label{s3}

To gain physical insights and accelerate the calculations, we will
adiabatically eliminate the excited states in the weak-driving limit in the
master equation, where the coupling strength between the ground and excited
states is much smaller than the energy-level broadening of the corresponding
excited state, i.e., $\Omega \equiv \sqrt{2/3}\Omega ^{\prime }\ll \Gamma _{%
\text{sd}}/2+\Gamma _{\text{pb}}$, which has been shown in many experiments.
Furthermore, when $\gamma _{\mathrm{e}}B_{x}\ll \gamma $, we can apply
linear response theory \cite{fetter2012quantum} and consider the effect of
the transverse field at the very end. Therefore, in this case we will drop
the Lindblad term $\mathcal{L}^{\left( 4\right) }\rho $ in the master
equation.

Adiabatic elimination in the master equation is common in quantum optics
when working with open systems \cite%
{Gardiner2004,PhysRevA.81.023816,PhysRevA.87.022110}. There are several ways
to accomplish adiabatic elimination. For example, one can utilize a
generating function, as is commonly done in the Fr\"{o}hlich transformation
\cite{cohen2006quantum}, but in the superspace, adiabatic elimination is
usually performed in the motion equations \cite%
{Gardiner2004,PhysRevA.81.023816,PhysRevA.87.022110}. Here, we apply the
latter to the alkali-metal-vapor atomic systems. Following the standard
procedure, we first define two projection operators $\mathcal{P}$ and $%
\mathcal{Q}=1-\mathcal{P}$, where $\mathcal{P}$ projects any given operators
in the Hilbert space or vectors in the superspace to the ground-state
subspace. For instance, when performing in the density matrix, $\mathcal{P}%
\rho $ gives%
\begin{equation}
\mathcal{P}\rho =\sum_{FMF^{\prime }M^{\prime }}\left\langle
s_{F}M\right\vert \rho \left\vert s_{F^{\prime }}M^{\prime }\right\rangle
\left\vert s_{F}M\right\rangle \left\langle s_{F^{\prime }}M^{\prime
}\right\vert .
\end{equation}

Next, we write the full master equation (\ref{me}) in the $\mathcal{P}$- and
$\mathcal{Q}$-spaces and adiabatically eliminate the $\mathcal{Q}$-space,
acquiring an effective master equation in the $\mathcal{P}$-space.\ For this
purpose, we separate the Lindblad operators in the full master equation (\ref%
{me})\ into two parts,%
\begin{equation}
\partial _{t}\rho =\left( \mathcal{L}_{0}+\mathcal{L}_{1}\right) \rho ,
\end{equation}%
where%
\begin{equation}
\mathcal{L}_{1}\rho \equiv -i\left[ H_{\text{lm}},\rho \right]
\end{equation}%
is the perturbation that couples the $\mathcal{P}$-space to the $\mathcal{Q}
$-space\, and $\mathcal{L}_{0}\rho =\left( \sum_{n=1}^{3}\mathcal{L}^{\left(
n\right) }-\mathcal{L}_{1}\right) \rho $ is the zeroth order term. Noting
that $\mathcal{P}+\mathcal{Q}=1$, $\mathcal{Q}${}$\mathcal{L}_{0}\mathcal{P}%
=0$, and $\mathcal{PL}_{0}\mathcal{P}=0$, we can write the density matrix's
evolution in the $\mathcal{P}$- and $\mathcal{Q}$-spaces respectively as%
\begin{equation}
\partial _{t}\mathcal{P}\rho =\mathcal{PL}_{0}\mathcal{P}\rho +\mathcal{PL}%
_{0}\mathcal{Q}\rho +\mathcal{PL}_{1}\mathcal{Q}\rho ,  \label{1}
\end{equation}%
\begin{equation}
\partial _{t}\mathcal{Q}\rho =\mathcal{Q}\text{{}}\mathcal{L}_{0}\mathcal{Q}%
\rho +\mathcal{Q}\text{{}}\mathcal{L}_{1}\mathcal{P}\rho +\mathcal{Q}\text{{}%
}\mathcal{L}_{1}\mathcal{Q}\rho .  \label{2}
\end{equation}%
To adiabatically eliminate the $\mathcal{Q}$-space, we solve $\mathcal{Q}%
\rho $\ from Eq.~(\ref{2})\ and substitute it in Eq.~(\ref{1}). The solution
for $\mathcal{Q}\rho $ in Eq.~(\ref{2}) is%
\begin{equation}
\mathcal{Q}\rho \left( t\right) =\int_{0}^{t}e^{\mathcal{Q}\left( \mathcal{L}%
_{0}+\mathcal{L}_{1}\right) \left( t-t^{\prime }\right) }\mathcal{Q}\text{{}}%
\mathcal{L}_{1}\mathcal{P}\rho \left( t^{\prime }\right) dt^{\prime },
\end{equation}%
where we assume an initial condition $\mathcal{Q}\rho \left( 0\right) =0$.
This assumption shows that the system is initially in the $\mathcal{P}$%
-space, which is reasonable, since before the interaction with the pump
laser, the steady state of the system is in the $\mathcal{P}$-space. Then,
substituting this solution of $\mathcal{Q}\rho \left( t\right) $ in Eq.~(\ref%
{1}), for the second order of $\mathcal{L}_{1}$, we acquire the density
matrix in the ground-state subspace,%
\begin{eqnarray}
\partial _{t}\mathcal{P}\rho \left( t\right) &\approx &\mathcal{PL}_{0}%
\mathcal{P}\rho \left( t\right) +\mathcal{PL}_{1}\int_{0}^{t}e^{\mathcal{Q}%
\text{ }\mathcal{L}_{0}\left( t-t^{\prime }\right) }\mathcal{Q}\text{{}}%
\mathcal{L}_{1}\mathcal{P}\rho \left( t\right) dt^{\prime }  \notag \\
&&+\mathcal{PL}_{0}\int_{0}^{t}e^{\mathcal{Q}\text{{}}\mathcal{L}_{0}\left(
t-t^{\prime }\right) }(1+  \notag \\
&&\int_{0}^{t-t^{\prime }}dt^{\prime \prime }e^{-\mathcal{Q}\text{{}}%
\mathcal{L}_{0}t^{\prime \prime }}\mathcal{Q}\text{{}}\mathcal{L}_{1}e^{%
\mathcal{Q}\text{{}}\mathcal{L}_{0}t^{\prime \prime }})\mathcal{Q}\text{{}}%
\mathcal{L}_{1}\mathcal{P}\rho \left( t\right) dt^{\prime },
\end{eqnarray}%
where we have applied the Born-Markov approximation \cite%
{Walls2008,Gardiner2004} to replace $\rho \left( t^{\prime }\right) $ by $%
\rho \left( t\right) $ in the integral and extend the upper limit $t$ in the
integration to $+\infty $. The Born-Markov approximation has been verified
in many quantum open systems \cite{Gardiner2004,Walls2008}, given that the
exponent $e^{\mathcal{L}_{0}t}$ decays on a time scale much smaller than
that of $\mathcal{P}\rho \left( t\right) $, which is the case in our system.

After straightforward calculations using the concrete expressions of $%
\mathcal{L}_{0}$ and $\mathcal{L}_{1}$, the effective master equation in the
ground-state subspace is
\begin{widetext}
\begin{eqnarray}
\partial _{t}\rho _{g} &=&-i\left[ H_{\text{hf}}+H_{\text{Zee}},\rho _{g}%
\right] +\gamma \left( \boldsymbol{S}\cdot \rho _{g}\boldsymbol{S-}\frac{1}{2%
}\left\{ \rho _{g},\boldsymbol{S}\cdot \boldsymbol{S}\right\} \right)
\notag \\
&&+\frac{1}{2}\gamma _{\text{se}}\left\langle S_{z}\right\rangle \left(
S_{+}\rho _{g}S_{-}-S_{-}\rho _{g}S_{+}\boldsymbol{+}\left\{ \rho
_{g},S_{z}\right\} \right) +\frac{1}{2}\gamma _{\text{se}}\left\langle
S_{+}\right\rangle \left( S_{-}\rho _{g}S_{z}-S_{z}\rho _{g}S_{-}\boldsymbol{%
+}\frac{1}{2}\left\{ \rho _{g},S_{-}\right\} \right) +\text{H.c.}  \notag \\
&&+\sum_{n=1}^{3}\sum_{FMF\prime M^{\prime }}\left[ \Gamma _{FMF\prime
M^{\prime }}^{(n)}J_{FF\prime M}^{\left( n\right) }\rho _{g}J_{FF\prime
M^{\prime }}^{\left( n\right) \dagger }\right] -\sum_{FMM^{\prime }}\left(
\Gamma _{FM}+\Gamma _{FM^{\prime }}^{\ast }\right) J_{FM}\rho
_{g}J_{FM^{\prime }}  \label{me1}
\end{eqnarray}%
\end{widetext}where $\ \rho_{g}=\mathcal{P}\rho \left( t\right) $ is the
density matrix in the ground-state subspace, the jump operators $J_{FF\prime
M}^{\left( n\right) }$ are

\begin{equation}
J_{FF\prime M}^{\left( 1\right) }=\left\vert F^{\prime },M+1\right\rangle
\left\langle FM\right\vert ,
\end{equation}%
\begin{equation}
J_{FF\prime M}^{\left( 2\right) }=\left\vert F^{\prime }M\right\rangle
\left\langle FM\right\vert ,
\end{equation}%
\begin{equation}
J_{FF\prime M}^{\left( 3\right) }=\left\vert F^{\prime },M+2\right\rangle
\left\langle FM\right\vert ,
\end{equation}%
the operator $J_{FM}$ is%
\begin{equation}
J_{FM}=\left\vert FM\right\rangle \left\langle FM\right\vert ,
\end{equation}%
and the pump-laser-induced relaxation rates are

\begin{widetext}

\begin{equation}
\Gamma _{FMF\prime M^{\prime }}^{(1)}=\frac{\Gamma _{\text{pb}}\Omega ^{2}}{%
\left\vert \tilde{\Delta} _{FF^{\prime }}\right\vert ^{2}}g_{2;FMFM^{\prime
}}g_{1;F^{\prime }M+1,F^{\prime }M^{\prime }+1},
\end{equation}%
\begin{equation}
\Gamma _{FMF\prime M^{\prime }}^{(2)}=g_{2;FMFM^{\prime }}g_{2;F^{\prime
}MF^{\prime }M^{\prime }}\sum_{F_{1}F_{2}}\frac{\Omega ^{2}\Gamma _{\text{pb}%
}}{\tilde{\Delta} _{FF_{1}}\tilde{\Delta} _{FF_{2}}^{\ast }}g_{1;F_{1}M+1,F_{2}M^{\prime
}+1}^{2},
\end{equation}%
\begin{equation}
\Gamma _{FMF\prime M^{\prime }}^{(3)}=g_{2;FMFM^{\prime }}g_{1;F^{\prime
}M+2,F^{\prime }M^{\prime }+2}\sum_{F_{1}F_{2}}\frac{\Omega ^{2}\Gamma _{%
\text{pb}}}{\tilde{\Delta} _{FF_{1}}\tilde{\Delta} _{FF_{2}}^{\ast }}g_{1;F_{1}M+1,F_{2}M^{%
\prime }+1}g_{2;F_{1}M+1,F_{2}M^{\prime }+1},
\end{equation}%
\begin{equation}
\Gamma _{FM}^{(4)}=\sum_{F^{\prime }}\frac{\Omega ^{2}}{\tilde{\Delta} _{FF^{\prime
}}}g_{2;FMFM}g_{1;F^{\prime }M+1,F^{\prime }M+1},
\end{equation}

\end{widetext}with the coefficients
\begin{equation}
g_{n,FMF\prime M^{\prime }}=CG_{F}\left( n,M\right) CG_{F^{\prime }}\left(
n,M^{\prime }\right)
\end{equation}%
and%
\begin{equation}
\tilde{\Delta}_{FF^{\prime }}=\Delta _{FF^{\prime }}-i\Gamma _{\text{pb}}.
\end{equation}%
Here, $CG_{F}\left( n,M\right) ,$ $n=1$, $2$, are the Clebsch-Gordan
coefficients, defined as%
\begin{equation}
CG_{F}\left( n,M\right) =\left\langle \left. I,M-\frac{\left( -\right) ^{n-1}%
}{2};\frac{1}{2},\frac{\left( -\right) ^{n-1}}{2}\right\vert FM\right\rangle
,
\end{equation}%
and $\Delta _{\text{aa}}=\Delta -\Delta _{\text{s}}$, $\Delta _{\text{ab}%
}=\Delta -\Delta _{\text{s}}-\Delta _{\text{p}}$, $\Delta _{\text{ba}%
}=\Delta $, and $\Delta _{\text{bb}}=\Delta -\Delta _{\text{p}}$ are the
energy differences. These energy differences $\Delta _{FF^{\prime }}$ are
the detunings between the pump laser's frequency and the transition
frequency between the ground state in the $F$ multiplet and the excited
state in the $F^{\prime }$ multiplet. We also define the corresponding
effective detunings $\tilde{\Delta}_{FF^{\prime }}$, given that each excited
state \textquotedblleft gains\textquotedblright\ an imaginary energy $%
-i\Gamma _{\text{pb}}$ representing its energy level broadening. Note that
we simplify the derivation of Eq.~(\ref{me1}) by assuming the hyperfine
splitting $\Delta _{\text{s}}$ is much larger than the (effective) decay
rates, which is the case for most alkali-vapor atomic magnetometers, such
that the coherence between the $a$ and $b$ multiplets, as well as the
contribution to the effective detuning $\tilde{\Delta}_{FF^{\prime }}$ from
the electronic spin relaxation and the Zeeman splitting, can be neglected.
Moreover, we assume that the spontaneous decay rate is much smaller than the
pressure broadening, $\Gamma _{\text{sd}}/2\ll \Gamma \equiv \Gamma _{\text{%
pb}}$, which occurs in atomic vapors with high pressure buffer gas. Thus the
spontaneous decay term can be neglected in the master equation. When the
condition $\Gamma _{\text{sd}}/2\ll \Gamma _{\text{pb}}$ is not met, we can
obtain a similar effective master equation in which the effective decay
rates $\Gamma _{FMF\prime M^{\prime }}^{(j=1,2,3)}$ and $\Gamma _{FM}^{4}$\
are slightly different.

The effective master equation (\ref{me1}) is valid in an extensive parameter
regime. In particular, when the energy splittings $\Delta _{\text{s}}$ and $%
\Delta _{\text{p}}\ $are both much smaller than the excited states' energy
broadening $\Gamma $, the master equation can be written in the compact form
\cite{PhysRevA.58.1412}%
\begin{eqnarray}
\partial _{t}\rho _{g} &=&\left( \mathcal{L}^{\left( 2\right) }+\mathcal{L}%
^{\left( 3\right) }\right) \rho _{g}  \notag \\
&&+\Gamma _{\mathrm{OP}}\left[ S_{+}\rho _{g}S_{-}+S_{z}\rho _{g}S_{z}+\frac{%
1}{2}\left\{ S_{z},\rho _{g}\right\} -\frac{3}{4}\rho _{g}\right]  \notag \\
&&-i\Delta _{\mathrm{LS}}\left[ S_{z},\rho _{g}\right] ,  \label{me2}
\end{eqnarray}%
where%
\begin{equation}
\Gamma _{\mathrm{OP}}=\frac{\eta ^{2}\Gamma }{\Gamma ^{2}+\Delta ^{2}}
\label{op}
\end{equation}%
is the optical pumping rate and%
\begin{equation}
\Delta _{\mathrm{LS}}=\frac{-\eta ^{2}\Delta }{\Gamma ^{2}+\Delta ^{2}}
\end{equation}%
is the light shift. This master equation (\ref{me2}) gives the Bloch
equations and the spin temperature distribution \cite%
{Walker1997,PhysRevA.58.1412}, where populations in states with the same
magnetic number $M$ are the same.

In an extensive parameter regime, including when the condition $\Delta _{%
\text{s,p}}\ll \Gamma _{\text{pb}}$ is not met, one can use the general
master equation (\ref{me1}) we have derived. It can be shown in (\ref{me1})
that when the coherence between the two multiplets $a\ $and $b$ in the spin
relaxation term are ignored, the diagonal elements of the density matrix $%
\rho _{g}$\ are decoupled from the diagonal ones. As a result, we obtain the
rate equations \cite{PhysRevA.60.3867}, i.e., the evolution of the diagonal
elements of the density matrix. In this case, only the diagonal terms are
non-vanishing in the steady-state solution to the master equation, i.e., the
polarization is along the $z$-direction and the mean values $\left\langle
S_{\pm }\right\rangle $ in $\mathcal{L}^{\left( 3\right) }$ are zero. This
reduces the number of coupled nonlinear equations from $4\left( 4I+2\right)
^{2}$ to $4I+2$, and speeds up the numerical calculation.

In the experiment \cite{Guo:19} with cesium atoms, whose nuclear spin is $%
7/2 $ and energy splittings are $\Delta _{\text{S}}=9.193$ GHz and $\Delta _{%
\text{P}}=1.168$ GHz, the atomic cell is cubic, with inner size $4\times
4\times 4\mathrm{mm}^{3}$, and is heated to $90$ Celsius. The power of the
pump beam is 700 $\mu $W, with right-handed circular polarization. The
magnetic field $B_{0}=0.1$G along the $z$-direction. Thus the atoms are
mostly pumped to states with negative magnetic numbers, and the polarization
is negative. This is equivalent to a left-handed circularly polarized pump
laser propagating antiparallel to the direction of the magnetic field. In
this case, we only need to change $B_{0}$ to $-B_{0}$ in the effective
master equation (\ref{me1}), while keeping the definition of the $z$%
-direction that defines the magnetic states. With the Rabi frequency $\Omega
=4.1$ MHz, spin exchange rate $\gamma _{\text{se}}=1.31 $ KHz, total spin
relaxation rate $\gamma =1.53$ KHz, and excited states' energy broadening $%
\Gamma =0.6$ GHz for the 100 torr nitrogen case, while $\gamma =1.65$vKHz, $%
\Gamma =4.2$ GHz for the 700 torr nitrogen case, we numerically solve the
master equation (\ref{me1}) for $\rho _{g}$ in the long term limit.

\begin{figure}[tbp]
\begin{center}
\includegraphics[width=0.8\linewidth]{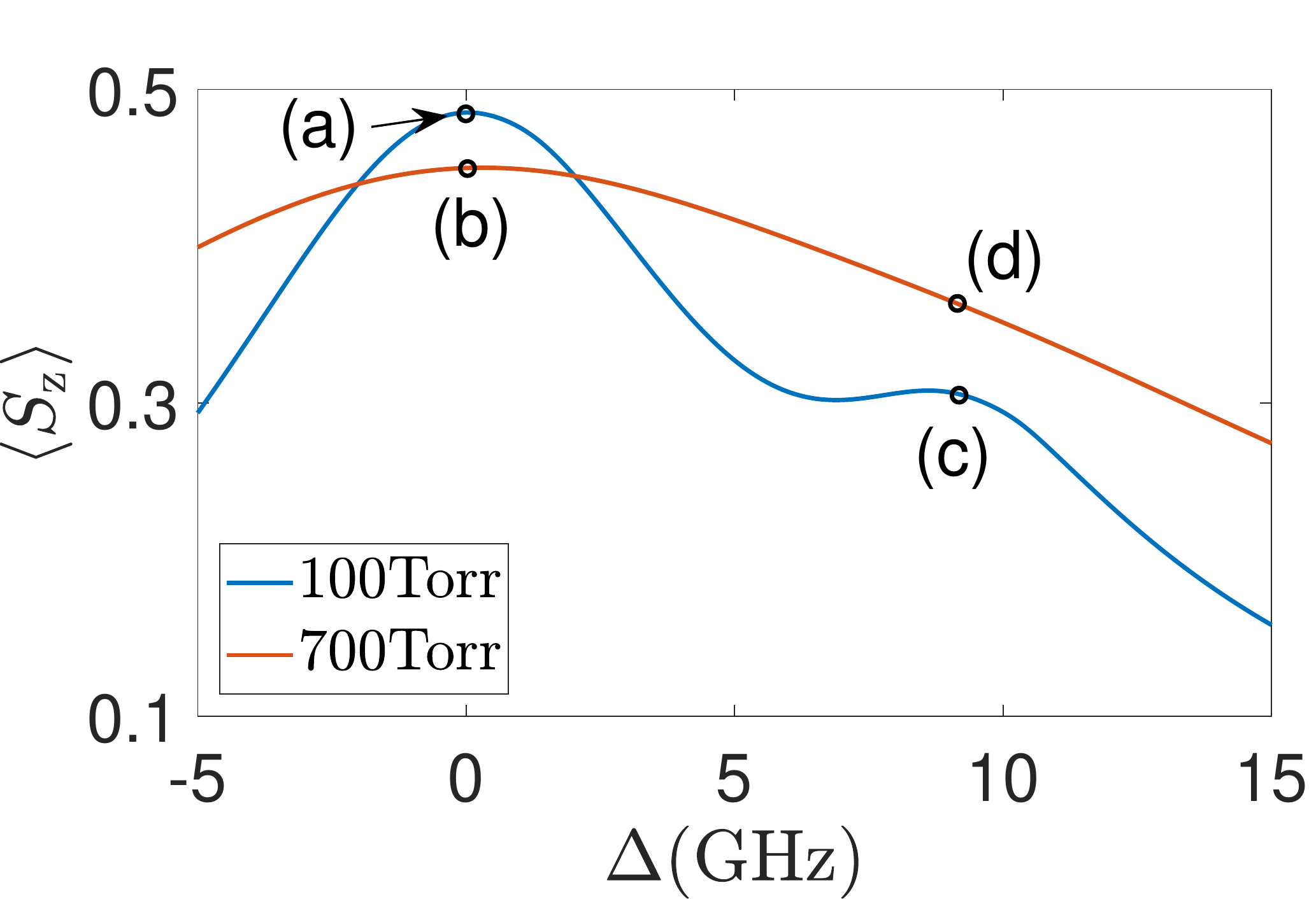}
\end{center}
\caption{Electron polarizations for cesium atoms in the steady-state
solution to the effective master equation in the ground state subspace, as
functions of the pump beam's detuning $\Delta$. For a smaller decay rate
(100 torr, blue line) of the excited state such that $\Gamma \ll \Delta _{%
\text{S}}$, there are two peaks in the electron's polarization, around two
resonant frequencies $\Delta=0$ (point (a)) and $\Delta _{\text{aa}}=0$
(point (c)). In the 700 torr case (red line), the decay rate $\Gamma$ is
comparable to $\Delta _{\text{S}}$. Therefore, the two polarization peaks
around the two frequencies ((b) and (d)) cannot be distinguished. The
populations in the ground states at these four points (a)-(d) are plotted in
Fig.~\protect\ref{fig3}.}
\label{fig2}
\end{figure}

\begin{figure}[tbp]
\begin{center}
\includegraphics[width=\columnwidth]{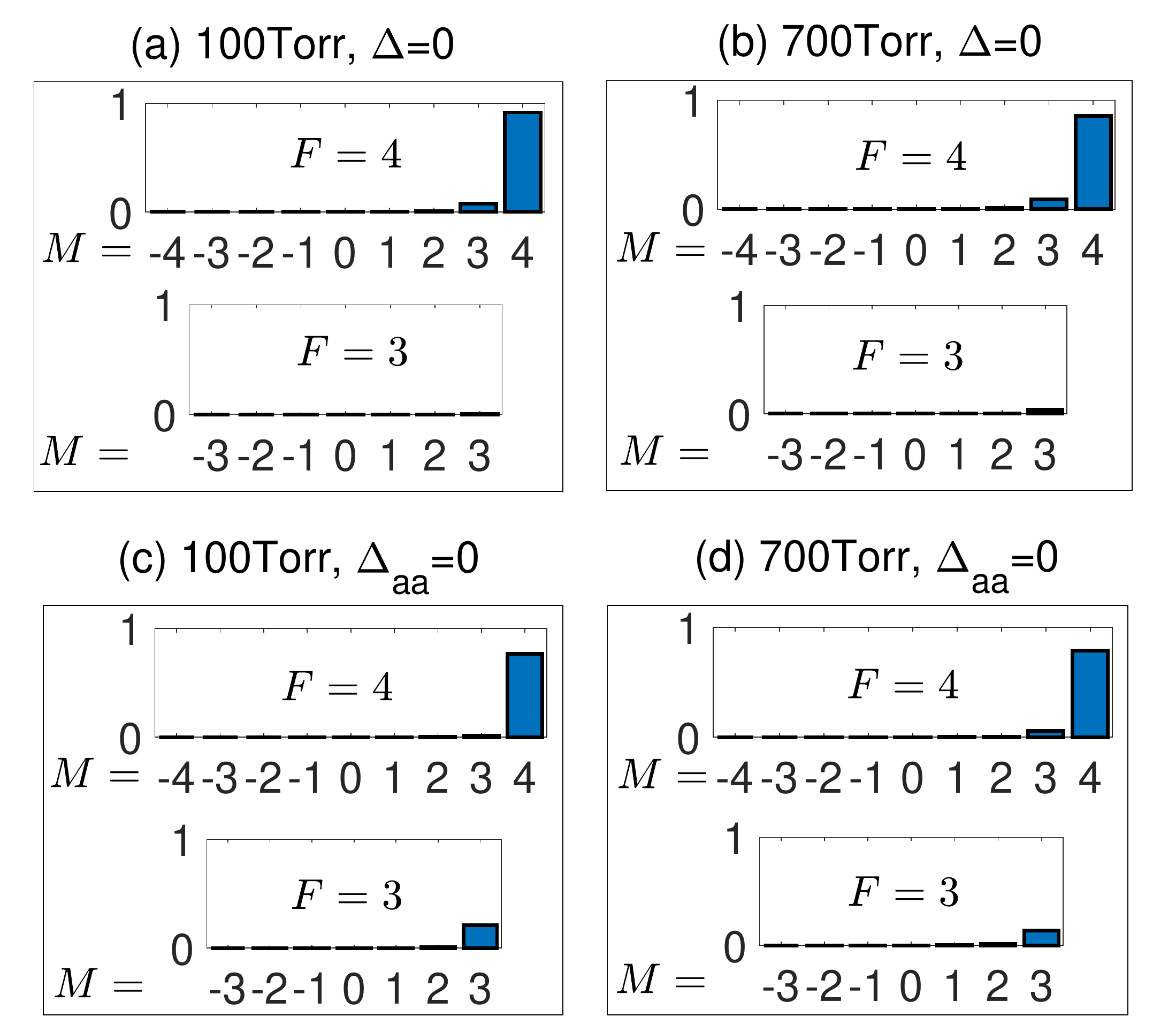}
\end{center}
\caption{Population distribution in the hyperfine states at two resonant
frequencies $\Delta=0$ ((a) and (b)) and $\Delta _{\text{aa}}=0$ ((c) and
(d)), for different energy level broadening $\Gamma$ of the excited state,
corresponding to the four points in Fig.~\protect\ref{fig2}. When the lower
ground states with $F=3$ are pumped ($\Delta=0$), the smaller the decay rate
$\Gamma$ the smaller the population in the $F=3$ states. However, when the
upper states with $F=4$ are pumped ($\Delta_{\text{aa}}=0$), a smaller $%
\Gamma$ gives a larger population in the $F=3$ states. Note that the
populations in states $\left\vert 4,m\right\rangle $ and$\ \left\vert
3,m\right\rangle $ are different, especially when $m=3$. Thus the spin
temperature distribution is not valid.}
\label{fig3}
\end{figure}

With the steady-state solution $\rho _{g}^{\left( 0\right) }$, where $\rho
_{g}^{\left( 0\right) }$ satisfies the effective master equation (\ref{me1})
and$\ \partial _{t}\rho _{g}^{\left( 0\right) }=0$, the electronic spin
polarization $\left\langle S_{z}\right\rangle $ as a function of the
detuning $\Delta $\ is plotted in Fig.~\ref{fig2}. For the 100 torr nitrogen
case, two peaks, corresponding to $\Delta \approx 0$ (marked by circle (a))
and $\Delta _{aa}\approx 0$ (marked by circle (c)), are shown in the
polarization curve, corresponding to two pump frequencies resonant with the
transition frequencies between the $a$/$b$ mutiplets and the excited states.
However, for the 700 torr nitrogen case, these two peaks (marked by circles
(b) and (d)) cannot be distinguished because of the large energy level
broadening $\Gamma $ of the excited states. Comparing the polarizations at
these four circles in Fig.~\ref{fig2}, we see that the polarization in (a)
is larger than in (b), while the polarization in (c) is smaller than in (d).
This is because the effective optical pumping rates $\Gamma _{FMF\prime
M^{\prime }}^{(j=1,3)}$ are inversely proportional to $\Gamma $. (We note
that the optical pumping process is generally complicated, as shown in Eq.~(%
\ref{me1}), and there does not exist a simple optical pumping rate, as shown
in Eq.~(\ref{op}).) In (a) and (b), $\Delta =0$ and the ground state with $%
F=3$ are more efficiently pumped and depleted, leaving the atoms mostly in
the $F=4$ states, which contribute more to the electrons' polarization. Thus
the larger the energy level broadening $\Gamma $ the smaller the
polarization. However, in (c) and (d), $\Delta _{aa}=0$ and the $F=4$ ground
states are more efficiently pumped, leaving the atoms populating the $F=4$
states less than in cases (a) and (b). Therefore, the larger $\Gamma $
causes more polarization. To verify this, we plotted the ground state
populations, i.e., the diagonal terms of the density matrix in Fig.~\ref%
{fig3}, for the four resonant cases (a)-(d) marked in Fig.~\ref{fig2}. Fig.~%
\ref{fig3} shows that the populations in the $F=4$ ground states are larger
in (a) and (d), compared with those in (b) and (c), respectively. Moreover,
in each case, the populations in states $\left\vert 4,m\right\rangle $ and$\
\left\vert 3,m\right\rangle $ are different, especially when $m=3$, which is
shown explicitly in the figures. Thus the spin temperature distribution \cite%
{Walker1997,PhysRevA.58.1412} is not valid.

Having solved the steady-state solution $\rho _{g}^{\left( 0\right) }$, we
will study the light shift and line width acquired from the linear response
\cite{fetter2012quantum} of the atoms to an oscillating transverse magnetic
field.

\section{Frequency-dependent light-shift reduction and light narrowing}

\label{s4} In the presence of the oscillating magnetic field $B_{x}\hat{e}%
_{x}\cos \omega t$ in the $x$-direction, where $\gamma _{\mathrm{e}}B_{%
\mathrm{x}}$ ($B_{\mathrm{x}}=3$nT in the experiment) is much smaller than
the decay rate $\gamma $ or $\Gamma _{FMF\prime M^{\prime }}^{(n)}$, the
master equation can be written as%
\begin{equation}
\partial _{t}\rho _{g}=\mathcal{\bar{L}}_{0}\rho _{g}+\mathcal{\bar{L}}%
_{1}\rho _{g}\text{,}
\end{equation}%
where%
\begin{eqnarray}
\mathcal{\bar{L}}_{0}\rho _{g} &=&-i\left[ H_{\text{hf}}+H_{\text{Zee}},\rho
_{g}\right]  \notag \\
&&+\gamma \left( \boldsymbol{S}\cdot \rho _{g}\boldsymbol{S-}\frac{1}{2}%
\left\{ \rho _{g},\boldsymbol{S}\cdot \boldsymbol{S}\right\} \right)  \notag
\\
&&+\frac{1}{2}\gamma _{\text{se}}\left\langle S_{z}\right\rangle \left(
S_{+}\rho _{g}S_{-}-S_{-}\rho _{g}S_{+}\boldsymbol{+}\left\{ \rho
_{g},S_{z}\right\} \right)  \notag \\
&&+\sum_{n=1}^{3}\sum_{FMF\prime M^{\prime }}\left[ \Gamma _{FMF\prime
M^{\prime }}^{(n)}J_{FF\prime M}^{\left( n\right) }\rho _{g}J_{FF\prime
M^{\prime }}^{\left( n\right) \dagger }\right]  \notag \\
&&-\sum_{FMM^{\prime }}\left( \Gamma _{FM}+\Gamma _{FM^{\prime }}^{\ast
}\right) J_{FM}\rho _{g}J_{FM^{\prime }}
\end{eqnarray}%
is the zero-order term, and%
\begin{eqnarray}
\mathcal{\bar{L}}_{1}\rho _{g} &=&\frac{1}{2}\gamma _{\text{se}}\left\langle
S_{+}\right\rangle \left( S_{-}\rho _{g}S_{z}-S_{z}\rho _{g}S_{-}\boldsymbol{%
+}\frac{1}{2}\left\{ \rho _{g},S_{-}\right\} \right)  \notag \\
&&+\text{H.c.}+\mathcal{L}^{\left( 4\right) }\rho _{g}
\end{eqnarray}%
is the first-order perturbation. Here, the zero-order Lindblad operator $%
\mathcal{\bar{L}}_{0}$ is different from the right side of Eq.~(\ref{me1})
regarding the terms containing $\left\langle S_{\pm }\right\rangle $, since $%
\left\langle S_{\pm }\right\rangle $ is zero from the zero-order solution $%
\rho _{g}^{\left( 0\right) }$. That is, the polarization in directions other
than the $z$-direction is induced by the magnetic field $B_{x}\hat{e}%
_{x}\cos \omega t$. As a result, the terms with $\left\langle S_{\pm
}\right\rangle $ are perturbations and it is in $\mathcal{\bar{L}}_{1}$
rather than in $\mathcal{\bar{L}}_{0}$.

To the first order of $\mathcal{\bar{L}}_{1}$, $\rho _{g}$ in the long-term
limit has three parts:%
\begin{equation}
\rho _{g}=\rho _{g}^{\left( 0\right) }+\rho _{g}^{\left( +\right)
}e^{i\omega t}+\rho _{g}^{\left( -\right) }e^{-i\omega t},
\end{equation}%
where $\rho _{g}^{\left( 0\right) }$ was obtained above by solving the
equation $\mathcal{\bar{L}}_{0}\rho _{g}^{\left( 0\right) }=0$; $\rho
_{g}^{\left( +\right) }$ is the positive-frequency part of $\rho _{g}$ that
fulfills%
\begin{equation}
\left( \mathcal{\bar{L}}_{0}-i\omega \right) \rho _{g}^{\left( +\right) }+%
\mathcal{\bar{L}}_{1}^{\left( +\right) }\rho _{g}^{\left( 0\right) }=0,
\end{equation}%
with its positive-frequency Lindblad operator $\mathcal{\bar{L}}_{1}^{\left(
+\right) }$ defined as%
\begin{eqnarray}
\mathcal{\bar{L}}_{1}^{\left( +\right) }\rho _{g}^{\left( 0\right) } &=&-%
\frac{i}{2}\gamma _{\mathrm{e}}B_{x}\left[ S_{x},\rho _{g}^{\left( 0\right) }%
\right] +\frac{1}{2}\gamma _{\text{se}}\text{Tr}\left[ S_{+}\rho
_{g}^{\left( +\right) }\right]  \notag \\
&&\times \left( S_{-}\rho _{g}^{\left( 0\right) }S_{z}-S_{z}\rho
_{g}^{\left( 0\right) }S_{-}\boldsymbol{+}\frac{1}{2}\left\{ \rho
_{g}^{\left( 0\right) },S_{-}\right\} \right)  \notag \\
&&+\text{H.c.;}
\end{eqnarray}%
and the negative-frequency part $\rho _{g}^{\left( -\right) }=\rho
_{g}^{\left( +\right) \dagger }$ to ensure the Hermitian of the density
matrix $\rho _{g}$. Note that $\mathcal{\bar{L}}_{1}^{\left( +\right) }$ is
dependent on $\rho _{g}^{\left( +\right) }$ through the mean value $%
\left\langle S_{+}\right\rangle $. Therefore, $\mathcal{\bar{L}}_{1}^{\left(
+\right) }\rho _{g}^{\left( 0\right) }$ can be decomposed to two parts: $%
\mathcal{\bar{L}}_{1}^{\left( +\right) }\rho _{g}^{\left( 0\right) }=%
\mathcal{\bar{L}}_{0}^{\left( +\right) }\rho _{g}^{\left( +\right) }+%
\mathcal{\bar{L}}_{1}^{\prime \left( +\right) }\rho _{g}^{\left( 0\right) }$%
, where
\begin{equation*}
\mathcal{\bar{L}}_{1}^{\prime \left( +\right) }\rho _{g}^{\left( 0\right) }=-%
\frac{i}{2}\gamma _{\mathrm{e}}B_{x}\left[ S_{x},\rho _{g}^{\left( 0\right) }%
\right],
\end{equation*}
and $\mathcal{\bar{L}}_{0}^{\left( +\right) }$ contains $\rho _{g}^{\left(
0\right) }$ but not $\rho _{g}^{\left( +\right) }$. As a result, the
solution of $\rho _{g}^{\left( +\right) }$ is%
\begin{equation}
\rho _{g}^{\left( +\right) }=-\left( \mathcal{\bar{L}}_{0}+\mathcal{\bar{L}}%
_{0}^{\left( +\right) }-i\omega \right) ^{-1}\mathcal{\bar{L}}_{1}^{\prime
\left( +\right) }\rho _{g}^{\left( 0\right) }.  \label{3}
\end{equation}%
Consequently, the electrons' polarization in the $x$-direction can be
written as%
\begin{equation}
\left\langle S_{x}\left( t\right) \right\rangle =\text{Re}\left\langle
S_{x}^{+}\right\rangle \cos \omega t-\text{Im}\left\langle
S_{x}^{+}\right\rangle \sin \omega t,
\end{equation}%
where $\left\langle S_{x}^{+}\right\rangle =2$Tr$\left[ S_{x}\rho
_{g}^{\left( +\right) }\right] $. Here, $\left\langle S_{x}^{+}\right\rangle
$ is a function of $\omega $. In the experiment, the measured Larmor
frequency $\omega _{\mathrm{L}}$ is determined by the zero-crossing $\omega
_{\mathrm{0}}$ of Re$\left\langle S_{x}^{+}\right\rangle $, and the line
width $w$ is defined as half the difference between frequencies
corresponding to the maximum and minimum$\ $of Re$\left\langle
S_{x}^{+}\right\rangle $.

As shown in Sec.~\ref{s3}, there are only diagonal terms in the steady-state
$\rho _{g}^{\left( 0\right) }$. Thus, in the superspace \cite%
{RevModPhys.44.169,happer2010optically,Gardiner2004}, $\mathcal{\bar{L}}%
_{1}^{\prime \left( +\right) }\rho _{g}^{\left( 0\right) }$ is a column
vector in the subspace $\left\{ \left\vert FM\right\rangle \left\langle
FM\pm 1\right\vert \right\} $, and $\mathcal{\bar{L}}_{0}+\mathcal{\bar{L}}%
_{0}^{\left( +\right) }$ is a matrix that does not couple this subspace $%
\left\{ \left\vert FM\right\rangle \left\langle FM\pm 1\right\vert \right\} $
to the others. In general, the zero-crossing $\omega _{\mathrm{0}}$ and the
line width $w$ are obtained by diagonalizing the matrix $\mathcal{\bar{L}}%
_{0}+\mathcal{\bar{L}}_{0}^{\left( +\right) }$, which can only be done
numerically. But to acquire an intuitive picture, we can analyze the
diagonal terms of $\mathcal{\bar{L}}_{0}+\mathcal{\bar{L}}_{0}^{\left(
+\right) }$.

When the Larmor frequency $\omega _{\mathrm{L}}$ is much larger than the
dissipation rates that contribute to the real parts of the eigenvalues of $%
\mathcal{\bar{L}}_{0}+\mathcal{\bar{L}}_{0}^{\left( +\right) }$, the
zeros-crossing $\omega _{\mathrm{0}}$ will be around $\pm \omega _{\mathrm{L}%
}$, the eigenvalues of $i\mathcal{L}^{\left( 2\right) }$ in the subspace $%
\left\{ \left\vert FM\right\rangle \left\langle FM\pm 1\right\vert \right\} $%
. Here, we focus on $\omega $ around the positive frequency $\omega _{%
\mathrm{L}}$, corresponding to the subspace $\left\{ \left\vert
aM\right\rangle \left\langle a,M+1\right\vert ,\left\vert bM\right\rangle
\left\langle b,M-1\right\vert \right\} $ (the coherence between states with
different $F$ has been ignored, for the same reason as in Sec.~\ref{s3}.).
Especially, when the atoms mainly populate the state $\left\vert
aa\right\rangle $, the most weighted diagonal element of $\mathcal{\bar{L}}%
_{0}+\mathcal{\bar{L}}_{0}^{\left( +\right) }$ is $i\tilde{\omega}-\tilde{%
\gamma}$ in the basis $\left\vert a,a-1\right\rangle \left\langle
aa\right\vert $, where the frequency%
\begin{equation}
\tilde{\omega}=\omega _{\mathrm{L}}+\frac{1}{2I+1}\frac{\Omega ^{2}\Delta
_{aa}}{\Gamma ^{2}+\Delta _{aa}^{2}}
\end{equation}%
and the line broadening%
\begin{eqnarray}
\tilde{\gamma} &=&\frac{\Omega ^{2}}{2I+1}\frac{\Gamma }{\Gamma ^{2}+\Delta
_{aa}^{2}}+\frac{I+1}{2I+1}\gamma  \notag \\
&&-\frac{1}{2I+1}\gamma _{ex}-\frac{2I}{2I+1}\gamma _{ex}\left\langle
S_{z}\right\rangle .
\end{eqnarray}%
In the light-narrowing regime with $\Delta \approx 0$, $\tilde{\omega}$ can
be approximated in the vicinity of this resonant frequency as%
\begin{equation}
\tilde{\omega}=\omega _{\mathrm{L}}-\frac{1}{2I+1}\frac{\Omega ^{2}\Delta _{%
\text{s}}}{\Gamma ^{2}+\Delta _{\text{s}}^{2}}+\delta \omega ,
\end{equation}%
where%
\begin{equation}
\delta \omega =\frac{1}{2I+1}\frac{\Omega ^{2}\Delta }{\Gamma ^{2}+\Delta _{%
\text{s}}^{2}}  \label{ls}
\end{equation}%
is the frequency-dependent light shift that leads to measurement inaccuracy
if the pump laser's frequency fluctuates. Because of the large hyperfine
splitting $\Delta _{\text{s}}$, the frequency-dependent light shift $\delta
\omega $ can be strongly reduced. Futhermore, for fully polarized atoms,
i.e., $\left\langle S_{z}\right\rangle =1/2$, the line width%
\begin{equation}
\tilde{\gamma}=\frac{\Omega ^{2}}{2I+1}\frac{\Gamma }{\Gamma ^{2}+\Delta
_{aa}^{2}}+\frac{I+1}{2I+1}\gamma _{\text{sd}},
\end{equation}%
and the spin-exchange relaxation does not contribute to the line width,
which makes perfect line narrowing \cite{PhysRevA.59.2078,PhysRevA.84.043416}
possible.

\begin{figure}[tbp]
\begin{center}
\includegraphics[width=\columnwidth]{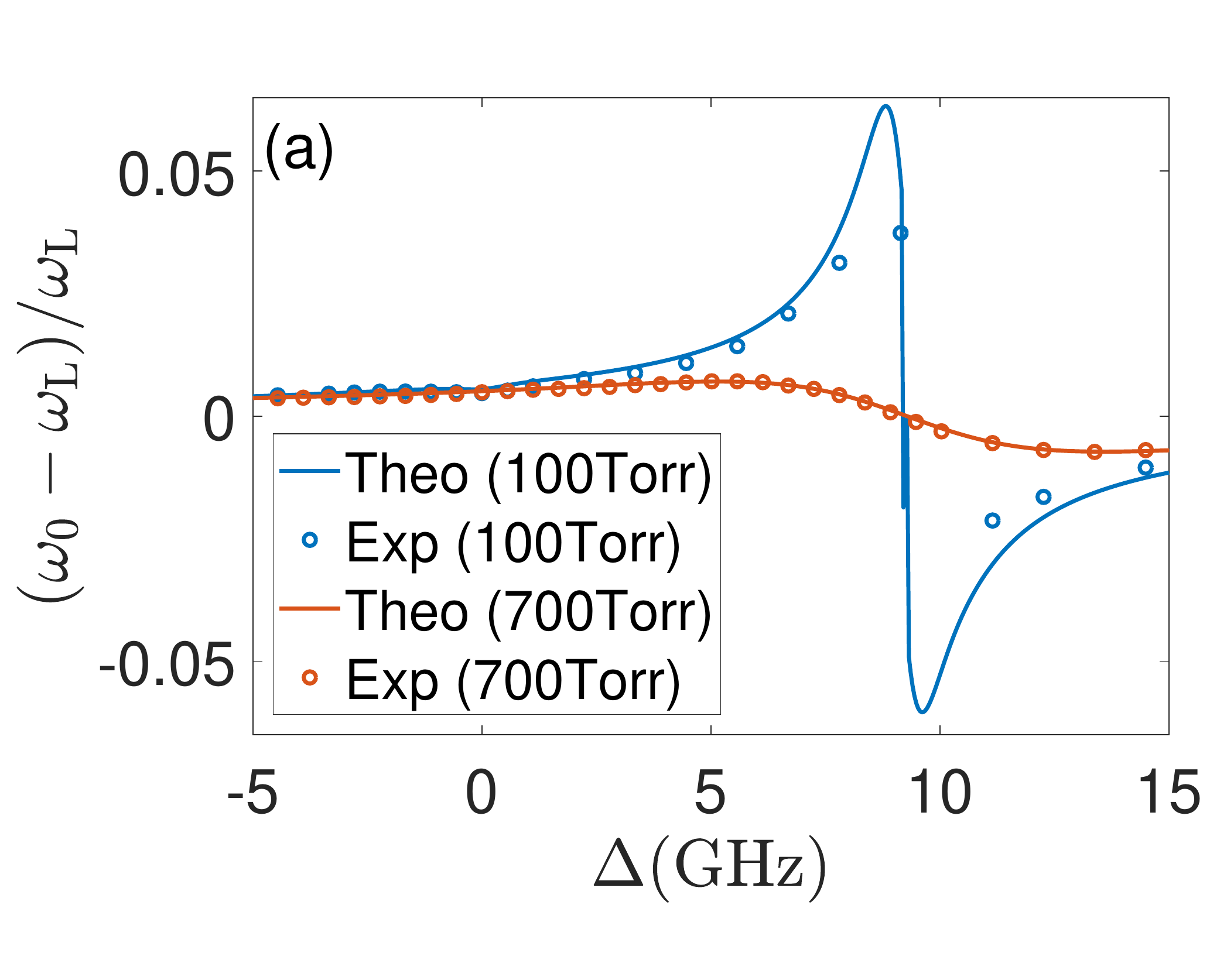} \includegraphics[width=%
\columnwidth]{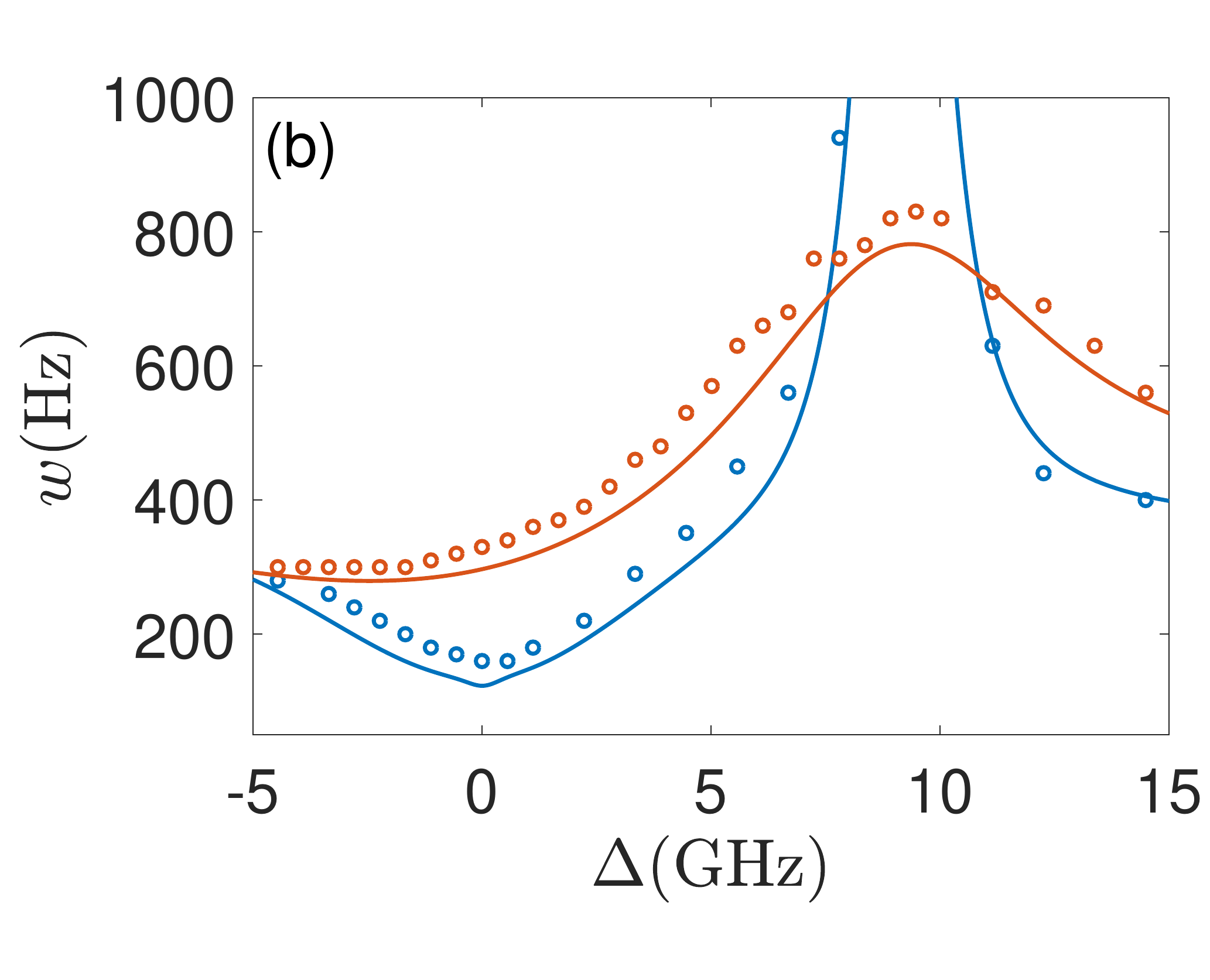}
\end{center}
\caption{Light shift (a) and line width (b) for the 100 torr and 700 torr
nitrogen cases as functions of the pump beam's detuning $\Delta$ (the legend
in (b) is the same as in (a)). The theoretical results are shown by the
lines, and the experimental data are shown by circles with corresponding
colors (the experimental data have been calibrated so that the light shift
at infinite detuning is zero). Both the theoretical and experimental results
show that the light shift's dependence on the pump laser's frequency is
largely reduced around the resonant point $\Delta$, especially compared to
the light shift around $\Delta_{\text{aa}}=0$. In this light-shift-reduced
regime, the line width is narrowed, as shown in (b). }
\label{fig4}
\end{figure}

The Lorentzian light shift $\delta \omega $ in Eq.~(\ref{ls}) is actually
the AC Stark shift. It gives an intuitive picture of why the light shift is
reduced in the light-narrowing regime. But, as shown in $\mathcal{\bar{L}}%
_{0}+\mathcal{\bar{L}}_{0}^{\left( +\right) }$ in Eq.~(\ref{3}), each pair
of adjacent magnetic levels has its own precession frequency (the imaginary
part of the diagonal terms of $\mathcal{\bar{L}}_{0}+\mathcal{\bar{L}}%
_{0}^{\left( +\right) }$), and they are all coupled (the non-zero
off-diagonal terms of $\mathcal{\bar{L}}_{0}+\mathcal{\bar{L}}_{0}^{\left(
+\right) }$). Thus the total light shift is generally not a single
Lorentzian or a sum of several Lorentzians. To obtain the exact result, we
numerically\ solve Eq.~(\ref{3}) and search for the zero-crossing $\omega _{%
\mathrm{0}}$ and the line width $w$. The numerical results, which are shown
in Fig.~\ref{fig4}, with the same parameters as in Fig.~\ref{fig2}, agree
well with the experimental data. For the light shift shown in Fig.~\ref{fig4}%
(a), in both the 100 and 700 torr cases, in the vicinity of the frequency $%
\Delta =\Delta _{\text{S}}$ ($\Delta _{\text{aa}}\approx 0$), where the $F=a$
ground states are pumped, the blue lines have two (100 torr) or one (700
torr) zero-crossings, corresponding to the resonant frequencies, and the
light shift changes much while the frequency varies. However, when $\Delta $
is around $0$, i.e., when the $F=b$ ground states are pumped, no
zero-crossing appears in the blue lines and the frequency-dependent light
shift is highly reduced. The line width shown in \ref{fig4}(b) has a dip
around the frequency $\Delta =0$. This is the light-narrowing effect.\ Note
that at a large detuning limit ($-5$ and $15$ GHz, for instance), the
light's effect tends to vanish. As a result, at infinite detunings, the
light shift goes to zero and the line width tends to be a constant,
independent of the pump beam's Rabi frequency $\Omega $, its detuning $%
\Delta $, or the excited states' decay rate $\Gamma $ \cite{Happer1977}.

\section{Conclusions and outlook}

\label{s5}

We have studied in detail the mechanism of the light shift and
light-narrowing effects in alkali-metal-vapor magnetometers. Starting from
the full master equation for the alkali atom's density matrix, we acquire
the effective master equation in the ground-state subspace by adiabatically
eliminating the excited states in the weak-driving limit. This effective
master equation cannot only save power and time for the numerical
calculations, but can reveal the intuitive picture of the
frequency-dependent light-shift reduction: in the light-narrowing regime,
the $F=b$ ground states are depleted by the pump laser, and the atoms mostly
populate the $F=a$ states. As a result, the light shift is reduced since the
pump beam's frequency is largely detuned from the transition frequency
between the most populated ground states ($F=a$) and the excited states. We
compare the theoretical results to the experimental data, and find they
agree for both the light shift and line width.

\begin{figure}[tbp]
\begin{center}
\includegraphics[width=\columnwidth]{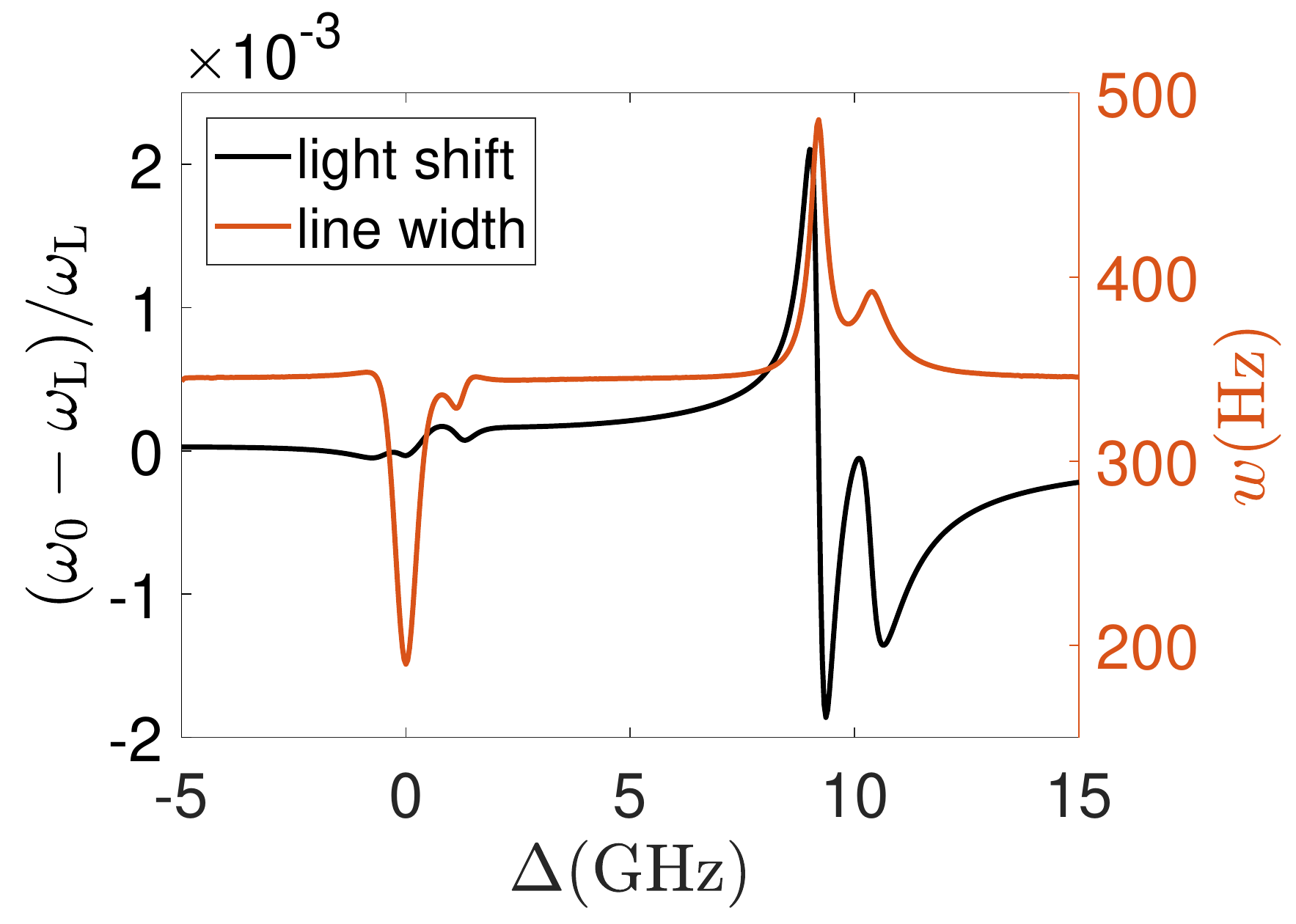}
\end{center}
\caption{Light shift (black line) and line width (red line) for $\Gamma=0.2$
GHz and $\Omega=0.5$ MHz, with other parameters the same as the 100 torr
nitrogen case in the cesium vapor experiment. With smaller $\Gamma$ and $%
\Omega$, more peaks and zero-crossings are shown in both the light shift and
line width.}
\label{fig5}
\end{figure}

We note that the effective master equation we have obtained is general and
is valid in an extensive parameter regime for alkali-vapor magnetometers.
Particularly, it can lead to the spin temperature distribution in the limit
that the hyperfine splittings in both the ground and excited states can be
ignored when the broadening of the excited states is much larger than them.
Since it consumes little time and power to solve this effective master
equation, one can use it to quickly explore a large parameter regime to
optimize the physical properties. For example, with a smaller decay rate $%
\Gamma =0.2$ GHz and Rabi frequency $\Omega =0.5$ MHz, while other
parameters are the same as in Fig.~\ref{fig2} for the 100 torr nitrogen
case, the light shift and line width are acquired and shown in Fig.~\ref%
{fig5}. Here, more peaks and zero-crossings can be distinguished,
corresponding to four resonant frequencies $\Delta _{\mathrm{FF^{\prime }}%
}=0 $ with $F$, $F^{\prime }=3,4$.

Besides the application shown in this paper, the effective master equation
is also applicable to many other topics, such as the study of heading errors
\cite{PhysRevA.79.023406,PhysRevLett.120.033202}, and light propagation in
an atomic vapor.

\acknowledgments This work was supported by the National Natural Science
Foundation of China grants 61473268, 61503353, and 61627806.

\bibliography{v6final.bbl}

\end{document}